\documentclass[twocolumn,prd]{revtex4-1}
\usepackage{amsmath}
\usepackage{amsfonts}
\usepackage{amssymb}
\usepackage{amsthm}
\usepackage{physics}
\usepackage{graphicx}
\usepackage{url}
\usepackage[pdfusetitle]{hyperref}
\usepackage{lmodern}
\usepackage{xcolor}

\usepackage[utf8]{inputenc}
\usepackage[T1]{fontenc}

\newcommand{\m}{\mu}

\newcommand{\n}{\nu}

\newcommand{\sg}{\sqrt{-g}}

\newcommand{\bee}{\begin{equation}}
\newcommand{\ee}{\end{equation}}
\newcommand{\beq}{\begin{equation*}}
\newcommand{\eeq}{\end{equation*}}
\newcommand{\baa}{\begin{equation}\begin{aligned}}
\newcommand{\ea}{\end{aligned}\end{equation}}
 
\newcommand{\qh}{\hat{q}}
\newcommand{\eh}{\hat{E}}
\newcommand{\lh}{\hat{\ell}}

\begin{document}
\title{Newton vs.\ Coulomb in AdS/CFT and the Weak Gravity Conjecture}
\author{Upamanyu Moitra}
\email{umoitra@ictp.it}
\affiliation{International Centre for Theoretical Physics (ICTP), Strada Costiera 11,  Trieste 34151,  Italy}

\begin{abstract}
We study (near-)circular orbits of charged particles in the background of charged black holes in asymptotically Anti-de Sitter (AdS) spacetimes of arbitrary dimensionality.  We calculate the energy and angular momentum of such particles in a large-radius limit.  This allows us to compute the anomalous dimension of the dual charged double-twist operators in a large-spin expansion,  making a prediction for the bootstrap analysis at large charge and spin.  We relate our result to the Weak Gravity Conjecture (WGC) for AdS black holes of all sizes.  We also discuss the relation of WGC with the existence of the innermost stable circular orbit (ISCO) in any dimension.
\end{abstract}

\maketitle

\section{Introduction}

Gravitation and electromagnetism are the two long-range forces present in nature. A quantitative understanding of the former began with Newton's law of gravitation, with Coulomb’s law having the corresponding distinction for the latter. We focus our attention on the interplay between the analogs of these force laws in this article. This interplay has deep implications for quantum gravity (QG). For instance, the Weak Gravity Conjecture (WGC) \cite{Arkani-Hamed:2006emk} posits that gravitation is weaker than electromagnetism in any consistent QG theory,  an observed fact of our universe. Isolating the space of consistent QG theories is the goal of the Swampland program \cite{Vafa:2005ui} in string theory.


The AdS/CFT correspondence \cite{Maldacena:1997re, *Gubser:1998bc, *Witten:1998qj} is one of the most powerful tools in understanding QG. It allows us to map QG questions in an asymptotically AdS$_{d+1}$ spacetime to those in the $d$-dimensional conformal field theory (CFT) on its conformal boundary and vice-versa. Depending on the context, one can calculate certain quantities in the bulk or the boundary side and translate it to the other side to extract physically meaningful information.  We will be studying the motion of charged particles in AdS to understand some aspects of the corresponding CFT.

Quite remarkably,  one can actually learn a great deal by studying such a simple system. There has been a recent focus on understanding some CFT operators with  bulk particles. In \cite{Berenstein:2020vlp},   the authors considered the orbit of a massive neutral particle in the AdS-Schwarzschild background and were able to reproduce features of the CFT obtained from a bootstrap analysis \cite{Fitzpatrick:2014vua}.

In this work, we extend these results in physically significant ways by considering charged particles in a charged black hole background. These considerations introduce several new physical ingredients. The presence of a bulk gauge symmetry translates to a global symmetry on the boundary. Charged black holes in the bulk correspond to heavy operators $\mathcal{O}_Q$ with a large U(1) global charge $Q$ on the boundary.  Charged probe particles in the bulk would be represented as light operators $\mathcal{O}_q$,  charged under the same global symmetry. This set-up is particularly interesting from the perspective of recent explorations of strongly coupled CFTs at large global charge \cite{Hellerman:2015nra};  see \cite{Gaume:2020bmp} for a review of related developments.  

In this work, we derive explicit analytic expressions for the energy and angular momentum of charged particles in large stable circular orbits in asymptotically AdS spacetimes of arbitrary dimensionality (the asymptotic AdS nature is crucial for the existence of stable orbits in five or higher bulk dimensions).  We also calculate the frequency of radial oscillations when slightly perturbed about the circular orbits. In the conformal bootstrap program, one is accorded a noteworthy simplification in the limit of large spin \cite{Alday:2007mf,  *Komargodski:2012ek,  *Alday:2015eya}.  Our calculation allows us to deduce the anomalous dimension of the double-twist operator $[\mathcal{O}_Q \mathcal{O}_q]_{n,J}$ in the dual CFT in a large-spin expansion.  Such operators, associated with bulk orbits,  have been related to light-cone bootstrap \cite{Kulaxizi:2018dxo,  *Karlsson:2019qfi,  *Li:2019zba, *Li:2020dqm} involving heavy and light operators,  for example in \cite{Dodelson:2022eiz}.  Our results thus make a new prediction for a similar bootstrap analysis at large spin and \emph{charge}.


One of the unique features of our set-up is the possibility of the existence of extremal black holes --- described by CFTs at zero temperature and a finite chemical potential. Such a set-up is useful also for describing condensed matter systems. One of the statements of WGC,  originally formulated for flat spacetime, is that any consistent QG theory must contain a particle with sufficiently high charge-to-mass ratio such that an extremal black hole is able to decay. The authors of \cite{Nakayama:2015hga} were the first to explore WGC in the context of AdS/CFT --- they proposed a bound related to the flat-space limit of AdS.  An extrapolation of this limit to large AdS black holes is, however,  non-unique and as emphasized in the recent review \cite{Harlow:2022gzl},  an appropriate formulation of WGC in AdS has remained an open problem so far.

Our results have a direct bearing on the WGC, in a manner which appears to be applicable for black holes of all sizes. The anomalous dimension that we find is a gauge-invariant characterization of the binding energy. Quite importantly, and in contrast with previous studies, the binding energy is sign-indefinite, reflecting the competition between Newtonian attraction and Coulomb repulsion (for like charges).  Demanding that this quantity be positive correlates exactly with the WGC bound in the flat-space limit. We use this observation to make comments about a possible formulation for WGC for large AdS black holes.

We further strengthen the connection between WGC and bulk orbit states by studying the innermost stable circular orbit (ISCO).  Curiously,  in dimensions five and above,  ISCOs do not exist in flat space.  One expects ISCOs to exist in the presence of even a small (negative) cosmological constant. Working in a large-AdS-radius perturbation theory,  we determine analytic expressions for the ISCO parameters for a charged particle and find that even in AdS,  ISCOs stop existing  precisely above the charge-to-mass ratio corresponding to WGC.  

Let us mention some relevant literature before proceeding.  A CFT binding energy formulation of WGC was proposed in \cite{Aharony:2021mpc}.  WGC in the context of AdS$_3$/CFT$_2$ was explored in \cite{Benjamin:2016fhe,  Montero:2016tif};  some other efforts to relate WGC and holography include \cite{Crisford:2017gsb,  Montero:2018fns}.

\section{Charged Particle Orbits in RN-AdS}

We consider Reissner-Nordstr\"{o}m-AdS (RN-AdS) black holes in $(d+1)$-dimensional spacetimes,  which arise as solutions to the Einstein-Maxwell-AdS action,
\baa
S =\int \dd[d+1]{x} \sg \pqty{R + \frac{d(d-1)}{L^2}  - F_{\m \n} F^{\m \n} }.  \label{aceinmax}
\ea
The asymptotically globally RN-AdS solution (with the AdS  ``radius'' $L$) is given by,
\baa
\dd{s}^2 = - f(r) \dd{t}^2 + \frac{\dd{r}^2}{f(r)} + r^2 \dd{\Omega}_{d-1}^2, \label{metricdef}
\ea
where $\dd{\Omega}_{d-1}^2$ is the line-element on the unit $S^{d-1}$ and,
\baa
f(r) = 1 + \frac{r^2}{L^2} -  \frac{2M}{r^{d-2}} + \frac{Q^2}{r^{2d-4}},  \label{redshif}
\ea
with the gauge field solution being given by,
\baa
A_\m \dd{x}^\m = A_t (r) \dd{t} =  - k_d \frac{Q}{r^{d-2}} \dd{t}, \label{gaugepot}
\ea
where $k_d \equiv \sqrt{(d-1)/2(d-2)}$. This is the most natural gauge choice for our problem.
The physical mass and charge are proportional to $M$ and $Q$ respectively.   We always consider $d \geq 3$.

For a fixed $L$,  there is a one-parameter family of extremal black holes,  whose mass and charge parameters can be expressed in terms of the extremal horizon radius $r_h$ ($f(r_h) = 0 = f'(r_h)$),
\baa
M_{\rm{ext}} &= r_h^d \qty(\frac{d-1}{(d-2) L^2}+\frac{1}{r_h^2} ),  \label{mext}\\
Q_{\rm{ext}}^2 &= r_h^{2(d-1)} \pqty{ \frac{d}{(d-2) L^2}+\frac{1}{r_h^2} }. 
\ea
We call an extremal AdS black hole large or small depending on the whether $r_h /L \gg 1$ or $\ll 1$.  Small AdS black holes correspond to the flat-space limit.  We can relate $r_h$ to the (extremal) chemical potential $\mu$,
\baa
r_h = L \sqrt{\frac{(d-2) [2 (d-2) \mu ^2-(d-1)]}{(d-1) d}}. \label{rhmu}
\ea
We have $|\mu| \geq k_d$ with $|\mu| \gg 1$ for large extremal black holes.

We now consider a probe particle of mass $m$ and charge $q$ in this geometry.  Instead of $q$,  we will find it convenient to work with the charge-to-mass ratio $\qh \equiv q/m$.  There are two conserved quantities associated with its motion,  arising from the invariance of the standard point-particle action under time translation and azimuthal translation: energy $\eh \equiv E/m$ and angular momentum $\lh \equiv \ell/m$ (each defined per unit mass).   {In contrast with the existing literature \cite{Pugliese:2011py},  our goal is to derive specific formulas for arbitrary-dimensional AdS/CFT.}

In terms of these quantities,  we have a particle moving in an effective 1-dimensional potential,
\baa
\dot{r}^2   + \underbrace{\pqty{1 + \frac{\lh^2}{r^2} } f(r) - (\eh + \qh A_t)^2}_{=V(r)} = 0\label{orbeq},
\ea
where $\dot{}$ refers to a derivative wrt the proper time of the particle.
Circular orbits of radius $r_c$ satisfy $V(r_c) = 0  = V'(r_c)$. The physically acceptable solutions for $\eh, \lh$ are,
\baa
\eh_c &= - \qh A_t +  f \frac{ \sqrt{\qh^2 r_c^2 A_t'^2-2 r_c f'+4 f}-\qh r_c A_t'}{2 f-r_c f'},  \\
\lh_c^2 &=  r_c^2  f \pqty{ \frac{ \sqrt{\qh^2 r_c^2 A_t'^2-2 r_c f'+4 f}-\qh r_c A_t'}{2 f-r_c f'} }^2 - r_c^2. \label{ellsq}
\ea
The functions and derivatives appearing above are evaluated at $r=r_c$.

In the asymptotically flat limit ($L \to \infty$) we obtain  for large values of the orbit radius $r_c$,
\baa
\eh \approx 1  +  \frac{d-4}{2}  \frac{M - k_d  \qh Q}{r_c^{d-2}}, \quad
\lh^2 \approx (d-2) \frac{M - k_d \qh Q}{r_c^{d-4}}.  \label{flaten}
\ea
For $d=3$,  we recover the familiar expressions.  Arbitrarily large circular orbits exist only for $M > k_d \qh Q$ and are unstable for $d \geq 4$; the orbits can be stable only in $d=3$. We seem to be living in exactly the  right dimensions. { This can be explained physically: the term in the effective potential representing the centrifugal barrier goes like $1/r^2$ in all dimensions;  both Newtonian and Coulomb potentials, on the other hand,  go like $1/r^{d-2}$.  It is thus easy to see that Newtonian attraction and centrifugal repulsion can play off against one another to create a stable minimum only in $d=3$ --- in higher dimensions,  the centrifugal barrier cannot win over gravitational attraction at small values of radius.}

{For a finite value of $L$,  however,  the story will be different since AdS acts as a confining box. }  We can consider different regimes while taking the large-$r_c$ approximation.  
It is clear from above that $M^{\frac{1}{d-2}} \ll r_c \ll \pqty{  M L^2}^{\frac{1}{d}}$ is an uninteresting domain ($\qh Q$  is at most the similar order of magnitude as $M$),  since the finite-$L$ corrections are not sufficient to guarantee a stable orbit.  An interesting regime is,
\baa
\frac{M}{r_c^{d-2}} \ll \frac{r_c^2}{L^2} \ll 1, \label{ineq1}
\ea
which is  appropriate for orbits barely feeling the effect of a non-zero cosmological constant. 
This regime is well-suited for  studying the effects of small extremal black holes.

In this approximation,
\baa
\lh^2 &= \frac{r_c^4}{L^2} \pqty{ 1 + L^2 (d-2) \frac{M -k_d \qh Q}{r_c^{d}    } },  \\
\label{angmom1}
\eh &\approx  1  + \frac{r_c^2}{L^2} +  \frac{d-4}{2}  \frac{M -k_d \qh Q}{r_c^{d-2}}.
\ea
We can invert the first relation to obtain $r_c$ as a perturbative expansion in large $\lh L$, which allows us express the energy in a coordinate-independent manner:
\baa
\eh &\approx  1 + \frac{\lh}{L}   - \frac{M - k_d \qh Q }{  (\lh L)^{d/2-1}   }   \label{energy1}.
\ea
The terms that are neglected are suppressed by powers of $1/\lh L$. We now consider the regime $r_ c \gg L,  M^{\frac{1}{d-2}}$,  which is necessary when we consider large black holes.  This regime is actually applicable for black holes of all sizes.  The perturbative expansion here is rather different (the following expressions are valid for $d \geq 5$),

\baa
\lh^2 &\approx  \frac{r_c^4}{L^2} \pqty{ 1 + \frac{P}{r_c^{d-2} } + (d-2)L^2 \frac{M - k_d \qh Q}{r_c^d} }, \\
\eh &\approx 1 + \frac{r_c^2}{L^2}  \pqty{ 1 + \frac{P}{2 r_c^{d-2} } +  \frac{(d-4)L^2 }{2}  \frac{M - k_d \qh Q}{r_c^{d}}   } \label{rhlin},
\ea
where $P \equiv d M - (d-2) k_d \qh Q$.  A similar exercise as before yields exactly the same result for the energy as \eqref{energy1}.  In fact,  for $r_c \gg L,  M^{\frac{1}{d-2}}$,  \eqref{energy1} holds true for all $d\geq 3$,  for large values of $\lh$, for black holes of all sizes.

We now consider the frequency of radial fluctuations about the circular orbit.  Starting from \eqref{orbeq}, staying at fixed angular momentum and taking care to normalize to boundary time \cite{Berenstein:2020vlp},   we have,
\baa
\omega_R = \frac{f(r_c)}{\eh_c + \qh A_t (r_c) } \sqrt{\frac12 V''(r_c) } .  \label{radfre}
\ea
In the regime $r_c \gg L,  M^{\frac{1}{d-2}}$,  we obtain from \eqref{radfre},
\baa
\omega_R L =  2 - d\frac{(d+2) M - \sqrt{\frac12 (d-2) (d-1)} \qh Q}{ 4 (\lh L)^{d/2 -1}}. \label{omr}
\ea

\section{CFT and WGC}

We are now ready to translate our results into CFT language.  A scalar field of mass $m$ and charge $q$ in the bulk is mapped to a boundary operator of dimension $\Delta_q = d/2 + \sqrt{m^2 L^2 + (d/2)^2}$.  We take $m \gg L^{-1}$ so that $\Delta_q \approx m L$.  By the AdS/CFT dictionary,  single-particle bulk states are mapped to single-trace CFT operators.  As in \cite{Berenstein:2020vlp, Dodelson:2022eiz},  we will use the map $\lh \to J (L/\Delta_q)$ and also define in the CFT $\qh \equiv q (L/ \Delta_q)$ (where $J$ and $q$ are the angular momentum and charge of the light operator respectively) to obtain the Regge trajectory for large $J$,
\baa
\Delta_{Q,q} (J) = \Delta_Q + \Delta_q + J + \gamma (M, Q, \qh , J), \label{regge}
\ea
where the first two terms correspond to ``rest energies'' of the black hole and probe particle,  $J$ is the spinning contribution to the energy and the anomalous dimension $\gamma$ corresponds to the binding energy,
\baa
\gamma (M, Q, \qh , J) \approx - \Delta_q \pqty{M - \sqrt{\frac{d-1}{2(d-2)}} \qh Q} \pqty{ \frac{\Delta_q}{J L^2} }^{\frac{d-2}{2}} . \label{gamma}
\ea
In the $Q\to 0$ limit,  our results reproduce, for example,  the results in \cite{Berenstein:2020vlp}. There is, however,  a crucial difference with their results. The anomalous dimension \eqref{gamma} can have either sign.  The possibility of $\gamma$ being positive alerts us to  a connection with WGC.  For a given black hole charge $Q  = Q_{\rm{ext}}$,  the minimum value of the mass is that of an extremal black hole $M = M_{\rm{ext}}$, see \eqref{mext}. Therefore,  the CFT anomalous dimension of a (like-charged) operator is non-negative if \footnote{This relation might strike some readers as the opposite of the usual WGC statement in which the inverse ratio,  i..e,  $(Q/M)_{\rm{ext}}$ appears.  We emphasize that this discussion is \emph{not} about self-repulsive particles, but about particles in  a black hole background.}
 \baa
\qh^2 \geq  \frac{2(d-2)}{d-1}  \frac{M_{\rm{ext}}^2}{Q_{\rm{ext}}^2}. \label{wgcbnd}
\ea
In the flat-space limit ($r_h \ll L$)  $M_{\rm{ext} } = r_h^{d-2} = Q_{\rm{ext} }$ and we obtain $\qh^2 \geq 2(d-2)/(d-1)$, which is precisely \footnote{In order to make connection with the WGC literature,  we can redefine the fields so that in the action \eqref{aceinmax},  $R$ and $F^2$ have coefficients $(2\kappa^2)^{-1}$ and $(-4e^2)^{-1}$ respectively.  If we reinstate these couplings, we shall have  $\qh^2 \to 2 e^2 \qh^2 / \kappa^2$.} the WGC bound \cite{Arkani-Hamed:2006emk,  Harlow:2022gzl}.  { While this bound does not involve $J$,   considering the large-$J$ regime might look unusual from the WGC perspective.  The usual formulation simply asserts the existence of a particle; if it exists,  the particle dynamics will be controlled by the laws discussed above.  Further studies of such regimes would be helpful in elucidating the WGC within AdS/CFT.} { It is worth emphasizing that the inequality \eqref{wgcbnd},  to begin with,  is independent of any statement of the WGC.  Given an AdS charged black hole and charged particles with parameters $(M, Q, m, q)$ \eqref{wgcbnd},  the inequality \eqref{wgcbnd} is simply the condition under which the gauge-invariant binding energy --- derived from the standard extrapolate dictionary --- is non-negative.  }

Since WGC requires the existence of particles meeting \eqref{wgcbnd} for small black holes,  one is tempted to extrapolate its validity in all regimes (this is not unnatural from the CFT perspective).  { In other words,  we want to see if the positive binding energy condition \eqref{wgcbnd} can be taken to be a possible statement of the WGC in a general manner.} We get $\mathcal{O}(1)$ bounds up to $r_h \sim L$.  In the limit of large extremal black holes $(r_h \gg L)$, \eqref{wgcbnd} is equivalent to,
\baa
\qh^2 \geq 2  \frac{d-1}{d} \pqty{ \frac{r_h}{L} }^2. \label{wgclbh}
\ea
This is a very large charge-to-mass ratio.  {While  such a particle orbiting a black hole is perfectly consistent in the gravitational theory,}  the existence of such states in a CFT would seem to be very unusual.   In a supersymmetric setting like $\mathcal{N} = 4$ SYM,  the mass is greater than the charge by the BPS inequality \footnote{{ In suitable units, we can write the BPS bound as $\Delta_q  \geq |q|$,  which gives an \emph{upper} bound on the  charge-to-mass ratio}}.  It is thus prudent to exercise caution in adopting \eqref{wgcbnd} as the defining statement of WGC,  in spite of the agreement in the flat space limit. However,  as emphasized in \cite{Dodelson:2022eiz}, the states we are discussing are narrow resonances and not necessarily states in the Hilbert space --- WGC particles could also be resonances.  It would be interesting to explore this question from the bootstrap point of view {\cite{Jafferis:2017zna}} especially in the non-supersymmetric context.

Extremal black hole geometries have a near-horizon AdS$_2$ factor.  As is well known in the holographic superconductivity literature \cite{Gubser:2008px,  *Hartnoll:2008vx}, a charged scalar can violate the AdS$_2$ Breitenlohner-Freedman bound \cite{Breitenlohner:1982jf} while simultaneously meeting the AdS$_{d+1}$ bound.  For large extremal black holes, we find the AdS$_2$ scaling dimension for a bulk scalar of mass $m$ and charge $q$ given by,
\baa
\Delta^{(2)}_q (\Delta^{(2)}_q - 1) = \frac{m^2 L^2 \pqty{ 1 - \frac12 \qh^2 } + \frac{L^2}{r_h^2} j (j + d -2)}{d(d-1)} ,
\ea
where $j$ is the spin quantum number. In the $s$-wave sector,  for instance,  $\Delta^{(2)}_q$ becomes complex well before \eqref{wgclbh} is met,  signifying an onset of horizon instability for $\mathcal{O}(1)$ values of $\qh$.    Unlike in the binding energy argument,  this instability is insensitive to the sign of $\qh$.   For large $j$,  one indeed needs a high value of $|\qh|$ for an instability. 

Nevertheless,  the principal lesson learned from WGC works quite well for large extremal black holes and $\mathcal{O}(1)$ values of $\qh$  \footnote{I am grateful to Cumrun Vafa for suggesting this argument.}.  We note from \eqref{mext} that in the large-charge regime,  $M \sim L^{-(d-2)/(d-1)} Q^{\frac{d}{d-1}}$.   WGC tells us that the effective field theory is valid when 
\baa
L^{-\frac{d-2}{d-1} } (Q + q)^{\frac{d}{d-1}} >L^{-\frac{d-2}{d-1} } Q^{\frac{d}{d-1}} + q.  \label{refwgc}
\ea
On reinstating the Planck length,  this means that the dimensionless charge must satisfy $Q > ( L/L_{\mathrm{Pl}})^{d-2}$.  We know from \eqref{mext} that  $Q \sim r_h^{d-1}/ L L_{\mathrm{Pl}}^{d-2}$ which meets this bound because for large black holes,  $r_h \gg L \gg L_{\mathrm{Pl}}$. 

From \eqref{omr},  we can find corrections to \eqref{regge} which are suppressed in $\Delta_q$, obtained by a Bohr-Sommerfeld--type quantization \cite{Berenstein:2020vlp,  Festuccia:2008zx},
\baa
\delta \Delta_{Q,q} (n) = 2n - n d\frac{(d+2) M - \sqrt{\frac12 (d-2) (d-1)} \qh Q}{ 4 (J L^2/\Delta_q )^{d/2 -1}},
\ea
where $n \geq 0$ is an integer. The second term above, contributing to the binding energy, is also sign-indefinite, but the sign changes at a different value of $\qh$ from \eqref{wgcbnd}.  For large black holes,  the inequality \eqref{wgclbh} changes by only an $\mathcal{O}(1)$ factor.  It is the lowest bound \eqref{wgclbh} that is the most important.

\section{ISCO and WGC}

Stable circular orbits cannot exist for arbitrary values of the radius. They stop existing at a radius $r=r_i$, at which the orbit becomes marginally stable, defined by the additional constraint $V''(r_i ) = 0$.  Such orbits are referred to as ISCO.

By numerical methods or otherwise,  we can straightforwardly demonstrate that for asymptotically flat RN black holes,  ISCOs can exist only for $d =3$,  even when the probe is charged.  This is related to our previous observation that in $d\geq 4$,  there exist no stable orbits in flat spacetimes.  It turns out that in $d=3$,  an ISCO for a positively charged extremal black hole stops existing precisely when $\qh >1$ (or $\qh < -1$ for $Q<0$) (see also \cite{Pugliese:2011py}).  This is the flat-space WGC bound for $d=3$.  We wish to connect WGC to ISCOs in higher dimensions as well. We immediately run against the aforementioned difficulty: there do not exist ISCOs in the first place.

The situation for $d\geq 4$ must change with a finite $L$.  In a large-$L$ perturbation theory,  we are able to derive the following formula for ISCO radius for $d \geq 5$,
\baa
r_i \approx \bqty{ \frac{(d-4)(d-2)}{4} \pqty{ M -k_d \qh Q  } L^2  }^{1/d}. \label{riexp}
\ea
When it exists,  $r_i$ diverges as $L \to \infty$,  explaining why we do not see them.  One can obtain the minimum value of angular momentum and the corresponding energy:
\baa
\lh_i^2 &\approx \frac{d}{d-4} \bqty{ \frac{(d-4)(d-2)}{4} \pqty{ M -k_d \qh Q  } }^{\frac{4}{d}} L^{\frac{8}{d} -2}, \\
\eh_i &\approx 1  + \frac{d}{d-2} \bqty{ \frac{(d-4)(d-2)}{4} \pqty{ M -k_d \qh Q  } }^{\frac{2}{d}} L^{\frac{4}{d} -2}.  \label{iscoel4}
\ea
Here,  however,  there is no clear separation between rotational and binding energies --- they contribute to the same order.  For sufficiently large values of $\qh Q$,  the quantity \eqref{riexp} becomes complex or negative,  which means that the ISCO stops existing.  For extremal black holes with $L \gg r_h$,  this happens precisely when the WGC bound \eqref{wgcbnd} is satisfied! The intuitive physical reasoning is that the Coulomb repulsion is strong enough to prevent the unstable and stable orbits from coalescing.  The connection between ISCO and WGC is thus intriguing and it is remarkable how AdS enters the story in an essential manner.

For $d=4$,  \eqref{riexp} is not valid and one instead has,
\baa
r_i \approx \bqty{  \frac{1}{4} L^2 \left(16 M^2-8 \sqrt{3} M \qh Q- (4-3 \qh^2) Q^2\right) }^{1/6},  \label{ri4ex}
\ea
which might suggest that ISCO exists for large $\qh$ as well.  By examining the energy and angular momentum (we have to insert \eqref{ri4ex} in the following expressions),
\baa
\eh \approx 1 + \frac{3r_i^2}{2L^2},  \quad \lh_i^2  \approx (2 M-\sqrt{3} q Q) + \frac{3r_i^4}{L^2},  \label{enl4exi}
\ea
we conclude that even in this case ISCOs stop existing above the WGC bound.  It is an interesting fact that in $d=4$,  even when $\qh = 0$,  ISCO radius \eqref{ri4ex} depends on $Q$,  which is not true for higher dimensions.

An analysis in the regime for large extremal AdS black holes is more difficult.  For \emph{uncharged} probes, we can self-consistently assume $r_i \gg r_h \gg L$ and obtain approximate expressions for the orbit parameters,
\baa
r_i \approx \bqty{  \frac{d(d+2)(d-1)}{4(d-2)} \frac{r_h^d}{L^2}  }^{\frac{1}{d-2} },\ea
\baa
 \eh \approx \lh / L \approx \sqrt{\frac{d+2}{d-2}} \bqty{\frac{d(d+2)(d-1)}{4(d-2)} }^{\frac{2}{d-2} }  \pqty{ \frac{r_h^2}{L^2} }^{\frac{d}{d-2}}.
\ea 
We thus have $\eh,  \lh \sim \mu^{2d/(d-2)}$ --- this parametric dependence on the chemical potential is the same as that observed for the temperature in the uncharged case \cite{Berenstein:2020vlp,  Festuccia:2008zx}.   This is to be expected since the chemical potential supplies the only scale in the extremal case.   Numerical investigations suggest that for charged probes around large black holes,  ISCO stops existing near the same value as \eqref{wgclbh}.

\section{Discussion}

It is worth emphasizing that all the results derived here involve global AdS as the asymptotic geometry.  Rather strikingly,  the analogues of ISCOs do not exist in the planar geometry.  This happens because, for instance,  there is no competing behavior between $r_h$ and $L$.  For the uncharged case,  the authors of \cite{Berenstein:2020vlp} interpreted this as an evidence for a non-perturbative effect in the boundary curvature,  which appears to be true for our scenario as well. 

The CFT states we considered could be interpreted as long-lived resonances, which could decay by some mechanism.  Being marginally stable,  ISCOs separate stable and unstable phases.  It is remarkable that the existence of ISCOs has a direct connection with WGC --- it would be interesting to explore the question of phase transitions in AdS/CFT and WGC.  Since extremal AdS black holes correspond to CFTs at zero temperature,  it would be illuminating to connect our picture with thermalization vis-\`{a}-vis the orbit states considered in this article.  See \cite{Dodelson:2022eiz} for comments pertaining to the uncharged case.  

In previous studies of binding energy in AdS,  it was the self-binding energy of particles in pure AdS spacetime that was considered \cite{Aharony:2021mpc,  Andriolo:2022hax}.  Our focus, instead,  has been on interaction between the particle and black holes of all sizes.  It would be interesting to explore the question of self-binding in such backgrounds,  in which case contact terms would play a very important role,  for small enough values of $\Delta_q$.  We can also gain substantial insight into these problems by considering specific examples of (holographic) CFTs. 

There is the prospect of extending this picture to other interesting geometries, including rotating black holes (for some very different approaches to WGC and related questions, see for instance,  \cite{Aalsma:2020duv, *Cuomo:2022kio}).  We can also introduce additional scalars.  It would also be worthwhile to consider in greater detail interactions with the black hole microstates,  which might be helpful in furthering our understanding of the dynamics of black holes in AdS/CFT.  We leave such exciting possibilities for future investigations.

\section*{Acknowledgements}
I thank Atish Dabholkar for useful discussions. I am particularly thankful to Cumrun Vafa for generously sharing his insights and for his comments on a draft of this article.
I thank the ICTP for support.

\bibliography{ncref}
\end{document}